\documentclass[10pt,a4paper,oneside]{report}
\usepackage[T1]{fontenc}
\usepackage{palatino,mathpazo}
\usepackage{amsmath,amsfonts,amssymb}
\usepackage{microtype}
\usepackage{graphicx}
\usepackage[table]{xcolor}
\usepackage{hyperref}
\hypersetup{
			colorlinks = true  , 
			urlcolor   = myBlue, 
			linkcolor  = myRed , 
			citecolor  = myRed   
			}
\usepackage{wrapfig}
\usepackage[small,sf,centerlast]{caption} 
\usepackage{booktabs}
\usepackage{geometry}
\usepackage{natbib}

\usepackage[scaled=1.03]{inconsolata} 
\usepackage{tgheros}                  
\def\MESA{\texttt{MESA}}

\def\teff{T_{\rm eff}}
\def\mast{M_\ast}
\def\mZAMS{M_{\mathrm{ZAMS}}}

\def\msol{M_\odot}
\def\lsol{L_\odot}

\def\llsol{L/L_\odot}

\def\nabad{\nabla_{\mathrm{ad}}}
\def\nabrad{\nabla_{\mathrm{rad}}}

\def\H1{^1\mathrm{H}}
\def\He3{^3\mathrm{He}}
\def\He4{^4\mathrm{He}}
\def\C12{^{12}\mathrm{C}}
\def\N14{^{14}\mathrm{N}}
\def\O16{^{16}\mathrm{O}}

\def\diff{{\mathrm d}}

\def\unity{ \hbox{1\kern-.23em l} }
\def\zero{ \hbox{0\kern-.23em |} }
\def\field{ \hbox{I\kern-.23em K} }
\newcommand{\afgtitle}[1]{{\LARGE #1}}
  
\newcommand{\afgauthor}[1]{{\large{\itshape #1}}}
\newcommand{\afgsection}[1]{{\scshape #1}} 
\newcommand{\afgsubsection}[1]{{\scshape #1}} 

\definecolor{Maroon}{RGB}{128,  0,  0}
\definecolor{myBlue}{RGB}{  8, 85,146}
\definecolor{myRed}{RGB} {138, 10, 11}

\pdfoutput=1
\begin{document}
 
\centerline{\afgtitle{On an Early~--~Post-AGB Instability}} 

\centerline{\textcolor{Maroon}{\noindent\rule{0.68\linewidth}{0.6pt}}}

\medskip
\centerline{\afgauthor{Alfred Gautschy}}
\centerline{\textit{CBmA 4410 Liestal, Switzerland}} 

\bigskip

\noindent\makebox[\textwidth][c]{
\begin{minipage}{0.8\linewidth}
{\small \noindent Dynamical stellar-evolution modeling through 
the AGB phase reveals that radial pulsations with very fast-growing
amplitudes develop if the luminosity to mass ratio of stars with
tenuous envelopes exceeds a critical limit. An instability 
going nonlinear already after a few cycles might qualify as a 
source of the \emph{superwind}~--~postulated to shed a
substantial part of a star's envelope over a very short time~--~
of hitherto persistently mysterious nature.}  
\end{minipage}}

\bigskip\bigskip

\centerline{\afgsection{1. INTRODUCTION}}

The evolution of stars along the asymptotic giant-branch (AGB) is
a realm of intricately interwoven threads of stellar microphysics
(advanced nuclear burning and hydrodynamic mixing and transport
processes), of macroscopic state changes due to thermal-pulse (ThP)
cycles, mass-loss processes, and all the consequences these stars
impose on the surrounding interstellar medium and the appearance of
the stellar populations of the hosting stellar systems
\citep{Herwig2005,VanWinckel2003}\footnote{
The references are mostly exemplary, they are meant
to be a guide to the extensive literature in this field.}.
Mass-loss in particular is an important actor because it potentially
exposes chemically modified deeper stellar layers and it channels the
intermediate- and low-mass stars into proto~--~white-dwarfs (WD) with
a mass distribution of characteristic shape.  Mass-loss along the AGB
is usually attributed to a combination of dust- and pulsation-driven
winds \citep{Freytag2023, Decin2021, Trabucchi2019}. The mass-loss
processes are far from easy to model quantitatively.  Even if they
were sufficiently well understood they might be beyond integration
into stellar-evolution computations due to conflicting timescales and
the computing power required.  Therefore, mass-loss processes are
typically added to evolution codes by means of observationally
motivated fit formulae: e.g.~\citep{Reimers1975} for mass-loss along
the first giant branch, and \citep{Bloecker1995} for mass-loss along
the AGB.

The top of the AGB is reached once the stellar envelopes, 
reduced by mass-loss, become thin enough so that
further evolution drives the stars away from the AGB on comparatively
short time scales. Depending on the rate of envelope-mass~--~shedding
and/or burning~--~AGB stars morph into hot, proto-WDs within
a few dozen to several tens of thousands of
years \citep{VanWinckel2003}. During the early post-AGB phase a 
short-lived, intensive dense wind~--~baptized as 
\emph{superwind}~--~was introduced to better reconcile
evolution modeling and the observed characteristics 
of planetary nebulae \citep{Renzini1981}. 
This superwind became a canonical part of post-AGB
evolution scenarios. Already at the inception, based on
rudimentary simulations of Mira pulsations, some 
sort of violent relaxation oscillations that built up on top of 
large-amplitude radial pulsations were conjectured as 
triggers of a superwind \citep{Wood1981}. The pulsations 
of AGB-stars as an efficient superwind source could, however, 
not be substantiated over all the years, so the basis of 
the superwind remains speculative.

This exposition reports on rapidly growing pulsations that were
encountered in some stellar-evolution computations around the terminal
AGB. In contrast to comparable findings of \citet{Wagenhuber1994} the
stellar-evolution results presented here were computed in dynamical
mode and the instabilities are found to mostly not be restricted to
particular phases within a ThP cycle so that it is conceivable that
such pulsations could eventually serve as the source of the enigmatic
superwind.

The report aims to attract the attention of specialists in AGB
astrophysics with access to sufficient computational power to
scrutinize and systematize the few \emph{proof-of-concept }results
presented here to stake the effects of such pulsations on the
properties of post-AGB stars, their circumstellar neighborhood,
and~--~most of all~--~to find out if such pulsations are compatible
with the extensive, intricate web of observational constraints
\citep[e.g.][]{Karakas2014, Herwig2005}.

\bigskip
\centerline{\afgsection{2. COMPUTATIONAL APPROACH}}

Stellar models referred to henceforth were computed with the
\MESA\,software instrument in the version close to what was described
in \citet{Paxton2019}. Canonically, most evolutionary computations are
made in quasi-hydrostatic equilibrium (QHE): Thermal imbalance is
accounted for in the evolution equations but the velocity term in the
momentum equation is neglected. However, most of the computations
underlying this study were carried out in \emph{dynamical} mode: The
acceleration term in the momentum equation was switched on by setting
\texttt{change\_v\_flag = .true.} and \texttt{new\_v\_flag = .true.}
in \MESA 's \texttt{inlist} file.  The Appendix to this exposition
contains relevant fragments of a representative \texttt{inlist} file;
it might be useful to reproduce the reported results. In this rough
exploratory study, sequences of model stars with $\mZAMS = 1.1, 1.5,
1.8, 2.0,$ and $3.0\,\msol$ were evolved from the ZAMS with
homogeneous compositions $X=0.7, Z=0.02$ (assumed prototypical of PopI
stars), and $X=0.757, Z=0.001$ to represent PopII stars. The model
stars' evolution was followed through central and shell helium burning
to the departure from the AGB and whenever possible to the WD cooling
regime.

\bigskip
\centerline{\afgsection{3. EVOLVING STARS THROUGH THE TERMINAL AGB PHASE}}

\begin{wrapfigure}{r}{0.50\textwidth}
	  \vspace{-20pt}
	  \begin{center}{
			\includegraphics[width=0.50\textwidth]{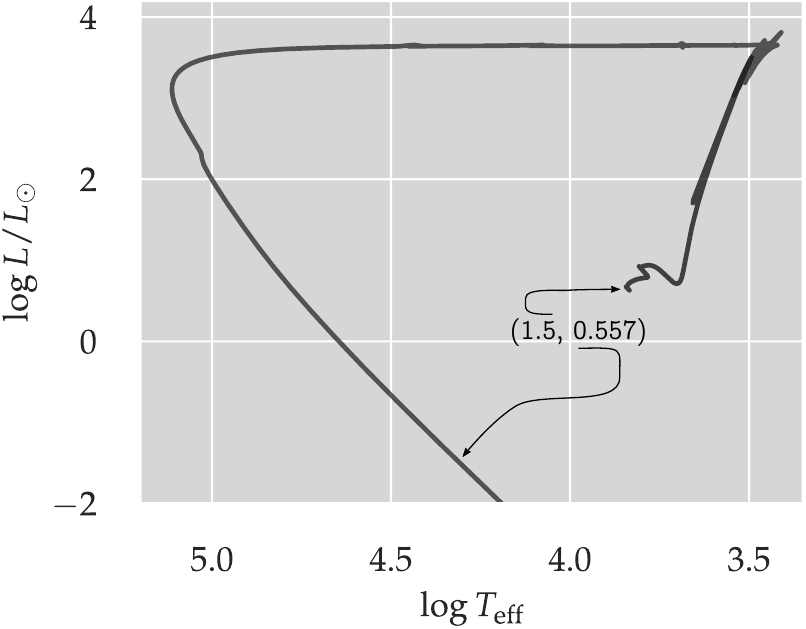}	
	  }\end{center}
	  \vspace{-15pt}
	  \caption{The evolutionary track from the ZAMS to 
	           advanced WD cooling of 
	           the $\mZAMS = 1.5\,\msol$ model star
	  		   that ended as a $0.557\,\msol$ DA white dwarf.
	  		   In this report, initial and final masses are 
	  		   coded by tuples as used in the figure.
	  	      }\label{fig:M015Z02_HRD}
	 \vspace{-20pt}	
\end{wrapfigure}
The mass-loss prescriptions adopted in the evolution computations 
boil down to quantifying the parameters that measure the strength of 
a Reimers- and a Bl\"ocker-type mass-loss recipe 
\citep[cf.][]{Paxton2011}. Initially, the canonical choices
\texttt{Reimers\_scaling\_factor = 0.1}  and   
\texttt{Blocker\_scaling\_factor = 0.2} were 
used to follow the stars' evolution as far as possible past 
the tip of the AGB. 

For easier restarts of evolutionary advanced sequences, a breakpoint
was set in the computations at around the onset of the
ThP-cycles. Additional model sequences (assuming hence the same
$\mZAMS$) could be restarted at these breakpoint-epochs with different
choices of the \texttt{Blocker\_scaling\_factor}.  Varying the
mass-loss efficiency caused the model stars to leave the AGB after
having lived through a different number of ThP cycles, arriving hence
at different final masses and also leaving the AGB at different phases
within a ThP cycle.  With respect to forcing stars off the AGB at
different evolutionary stages, this approach reminds of fishing in
muddy waters, which it actually is. Nonetheless, the outcome proofed
sufficiently useful for this exploratory study.

Forcing the model stars to leave the AGB after different numbers of
ThP cycles produced varying proto-WD masses. Leaving the AGB at
different phases in their ThP cycles meant that they started departing
the AGB at different luminosities and effective temperatures and hence
had different $L/M$ ratios.  It is important to emphasize that varying
the \texttt{Blocker\_scaling\_factor} is not physically motivated; it
is just a convenient numerical \emph{dial }to turn to enforce
departure from the AGB at will to observe the effect it has on the
stability of the model stars.

Figure~\ref{fig:M015Z02_HRD} shows the evolutionary track on the HR
plane of an $\mZAMS = 1.5\,\msol$ star model with an initially
homogeneous ZAMS $X=0.7,Z=0.02$ composition.  As the model evolved up
the AGB, it passed through four ThP cycles before the remaining
envelope mass got too thin and the star terminated its ascent. The
model left the AGB with a remaining mass of $0.557\,\msol$. The
essentially horizontal post-AGB locus shows some minor hiccups but
otherwise quietly enters the terminal WD cooling phase. The apparent
noise along the almost horizontal post-AGB track are, at closer
examination, not signs of poor convergence of the models.  Rather,
they are phases in which small oscillations develop, but which quickly
subside again.

\begin{wrapfigure}{l}{0.58\textwidth}
	  \vspace{-20pt}
	  \begin{center}{
	   \includegraphics[width=0.56\textwidth]{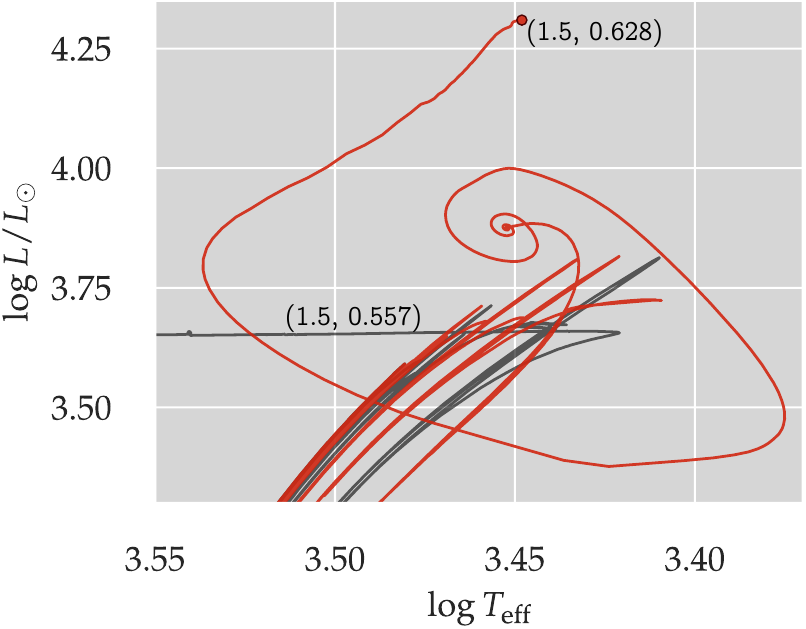}
	  }\end{center}
	  \vspace{-15pt}	  	
	  \caption{Zoom-in to the top of the AGB. The black locus is the
	           same $1.5\,\msol, Z=0.02$ track as shown 
	           in Fig.~\ref{fig:M015Z02_HRD}. Superimposed in red is
	           the track from the $(1.5, 0.628)$ sequence that was 
	           subjected to a different Bl\"ocker-type mass-loss rate 
	           during its thermally pulsing AGB phase. 
	  	      }\label{fig:M015Z02_AGB}	
	  \vspace{-5pt}	
\end{wrapfigure}
Very late during the AGB evolution, once the model stars' envelopes
get thin enough to initiate the evolution away from the AGB,
pulsations with large growth rates were found to develop under
favorable circumstances.  Figure~\ref{fig:M015Z02_AGB} illustrates how
this looks like in the case of the $(1.5, 0.628)$ sequence for which,
after seven ThP cycles, a high-luminosity departure from the AGB was
found for the \texttt{Blocker\_scaling\_factor = 0.2} choice.  The
rapidly growing spiral traces the pulsational instability on the HR
plane. Facing such an instability, the evolutionary computations were
found to terminate either due to too short timesteps or failing
convergence of \MESA\,once the amplitude of the pulsation grows too
large (the termination epoch is marked by the red dot in
Fig.~\ref{fig:M015Z02_AGB}). The pulsationally stable $(1.5,0.557)$
sequence (grey locus) was computed with {\tt Blocker\_scaling\_factor
  = 0.4}. As mentioned before, the respective star evolved through
four ThP cycles before it left the AGB to transit to the WD cooling
branch without any signs of an unfolding oscillatory instability.

\newpage

\centerline{\afgsubsection{3.1. THE PULSATIONAL INSTABILITY}}

To gather experience on the frequency of the occurrence of pulsations
of the kind stumbled over in the $\mZAMS = 1.5\,\msol$ case reported
in the last section, additional model sequences were computed with
different choices of {\tt Blocker\_scaling\_factor} in the hope of
discovering further cases that develop pulsations.

\begin{table}[ht!]
{
\centering
\caption{$Z=0.02$ model sequences that developed pulsations. 
        The pulsations of the
		highlighted row are representative and discussed
		vicariously in the text.} 
\label{Table:PulsationProperties}
\begin{tabular}[t]{lcrrc}\toprule
	 \centering
	  ($M_{\mathrm{ZAMS}},M_{\mathrm{final}}$) in s.u. 
	               &  $\log \llsol$ & P / d   & $\#$ of puls. cycles & {\tt Blocker\_scaling\_factor} \\ 
	 \midrule
	  (1.5, 0.628) & 3.87  &  1004   &   3 &  0.2 \\ \addlinespace
	  (1.8, 0.605) & 3.80  &   436   &   7 &  0.1 \\ \addlinespace 
	  \rowcolor{gray!20}(2.0, 0.570) & 3.74 &  98 &  11 & 0.4 \\ \addlinespace  
	  (2.0, 0.650) & 3.85  &  779    &   3 &  0.1 \\ \addlinespace  
	  (3.0, 0.781) & 3.97  &  1288   &   3 &  0.1 \\ \addlinespace  
	  (3.0, 0.641) & 3.86  &  719    &   4 &  1.0 \\ 
	 \bottomrule
\end{tabular}\par
} 
\end{table}

\begin{wrapfigure}{r}{0.55\textwidth}
	  \vspace{-20pt}
	  \begin{center}{
			\includegraphics[width=0.54\textwidth]{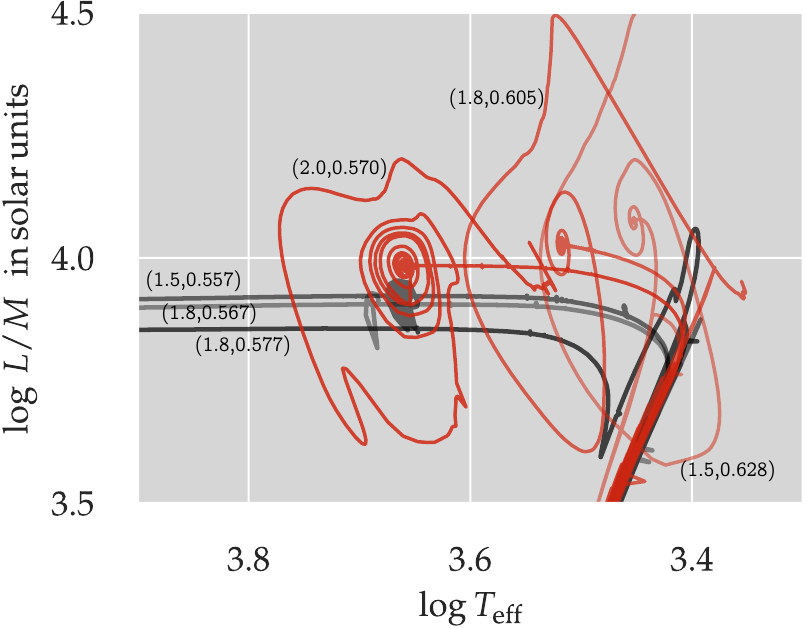}	
	  }\end{center}
	  \vspace{-15pt}
	  \caption{ Selected evolutionary tracks of the 
	            PopI ($Z = 0.02$) model sequences on the
	         	$\teff - L/M$ plane hint at tracks reaching
	         	sufficiently large $L/M$ ratios to be prone
	         	to pulsations. 
	  	      }\label{fig:M020Z02_LtoM}
	 \vspace{-20pt}	
\end{wrapfigure}
Table~\ref{Table:PulsationProperties} collects useful properties of
those model sequences, characterized by the respective
(initial-,final-mass) tuples in the first column, that developed
radial pulsations at the end of their AGB evolution.  The magnitude of
the only parameter that was changed in the \texttt{inlist} for a given
initial mass, {\tt Blocker\_scaling\_factor}, is listed in column
five.  The number of pulsation cycles that could be followed with
\MESA\,before it stalled is collected in the fourth column.  Column
two lists the luminosity at which the pulsations started to grow. The
third column finally contains the pulsation period in days, derived as
the average over the computable pulsation cycles.

Columns two and three of Table~\ref{Table:PulsationProperties} reveal
that the pulsation period and the star's luminosity are
correlated. The brighter the star, the longer is the period.  Mapping
the data from the table onto the $PL$-relations in
\citet{Trabucchi2019} shows, however, that the results reported here
do not match a single branch of their relations. Compared with LPV
pulsations discussed in the literature, the pulsations from
Table~\ref{Table:PulsationProperties} are generally overluminous at a
given period.

As it is evident from Table~\ref{Table:PulsationProperties},
pulsations develop in all sequences with $\mZAMS > 1.5 \msol$.  None
of the mass-loss choices adopted in the $\mZAMS = 1.1 \msol$ case
allowed the model stars to pulsate around the termination of their AGB
evolution. Evidently, the higher the starting mass of the model stars
is, the easier it seems for them to pick up rapidly-growing
pulsations.  The collection of evolutionary tracks devoid of
pulsations (grey) and respective ones (i.e. same $\mZAMS$) with
terminal pulsations (red) on the $\log \teff - \log L/M$ plane of
Fig.~\ref{fig:M020Z02_LtoM} hints at the $L/M$ ratio to have to exceed
a critical level for pulsations to develop. To keep
Fig.~\ref{fig:M020Z02_LtoM} reasonably readable, only tracks of $1.5,
1.8,$ and $2.0\,\msol$ sequences are included.

The horizontal post-AGB tracks of the stable (grey) evolutionary
sequences in Fig.~\ref{fig:M020Z02_LtoM} all display a few noisy
phases, some even with non-negligible amplitudes (such as along (1.5,
0.557) and (1.8, 0.567) tracks).  Such fidgeting is not a sign of
numerical convergence problems.  Closer inspection of the
larger-amplitude wiggles shows that these are associated with phases
of small-amplitude pulsations that failed to overcome the first growth
regime.  The iterative process to compute a model star usually
generates structural perturbations that may serve as seeds for such
instabilities to grow. Along the grey loci, perturbations can
apparently persist for a short while but eventually they damp out
again. In contrast, along the red evolutionary tracks, pulsations
unfold with large growth rates so that they reach the nonlinear regime
within a few pulsation cycles.

\bigskip
\centerline{\afgsubsection{3.2. POPULATION II STARS}}

\begin{wrapfigure}{l}{0.55\textwidth}
	  \vspace{-20pt}
	  \begin{center}{\includegraphics[width=0.54\textwidth]
			         {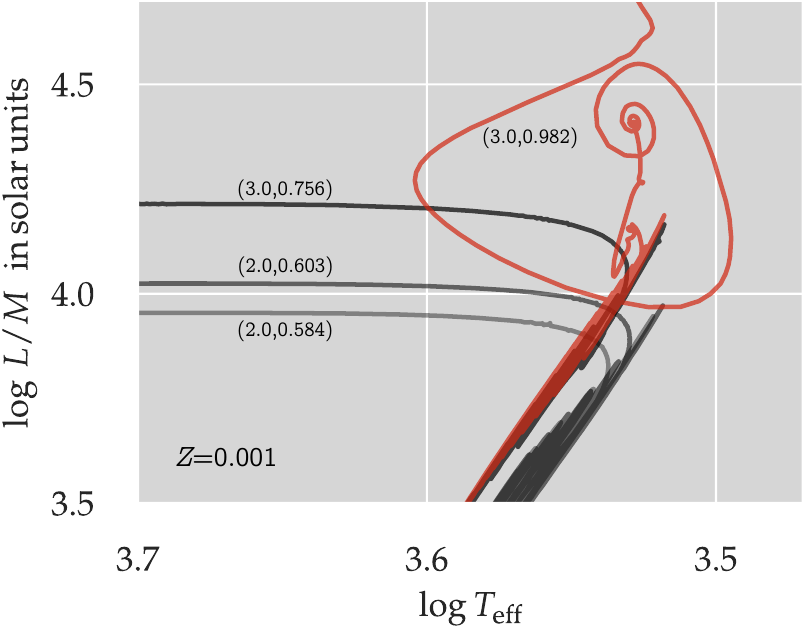}	
	                }
	  \end{center}
	  \vspace{-15pt}
	  \caption{ The red locus traces on the $\log\teff - \log L/M$ plane
                the final AGB evolution and the pulsation encountered in 
	            the (3.0,0.982) PopII sequence. The stable evolutionary
	            tracks of the $\mZAMS = 2.0$ and $3.0\,\msol$ choices
	            are plotted in grey. 
	  	      }\label{fig:Z001_pulsations}
	 \vspace{-20pt}	
\end{wrapfigure}
The results presented in Sect.~3.1 are appropriate for PopI-like stars
only.  In the astrophysical context of AGB evolution, the behavior of
more metal-poor stars is relevant and pulsations are usually sensitive
to the chemical composition of the stellar matter. Hence, additional
model sequences were computed that started with a PopII-like
homogeneous ZAMS composition of $X = 0.758, Z = 0.001$.  Metal-poor
stars are known to be bluer than their brethren with higher
heavy-element abundances. The PopII evolutionary tracks up the AGB are
comparatively hotter, and the configurations are therefore more
compact than respective PopI model stars.

As Table~\ref{Table:PulsationPropertiesPopII}~--~the
table-of-one-row~--~already hints at, in PopII stars early~--~post-AGB
pulsations are harder to find than in PopI models. In case of the
adopted PopII-like composition, stars with $\mZAMS \lesssim 3\,\msol$
fail to develop the sought pulsations. Currently, only one case can be
reported: The $(3.0, 0.982)$ sequence displayed on the $\log \teff -
\log L/M$ plane of Fig.~\ref{fig:Z001_pulsations}.  The model departed
from the AGB at around $\log L/\lsol = 4.4$, which is considerably
higher than what is seen in respective PopI sequences. The pulsation
period, averaged over the three computable cycles, was accordingly
long with $1097 $~days. The motivation to choose a low value of the
Bl\"ocker-type mass-loss was to force to star to stay longer, i.e. to
live through more ThPs that take it further up along the AGB with each
cycle.  The $16$~pulses the $(3.0, 0.982)$ sequence went through were
eventually enough to build up a sufficiently tenuous outer envelope in
which a pulsation developed.  Figure~\ref{fig:Z001_pulsations}
suggests that, compared with the $Z = 0.02$ sequences, PopII stars
require a larger $L/M$ ratio to host pulsations.

\begin{table}[ht!]
{
\centering
\caption{The $Z=0.001$ model sequence that developed a 
         terminal pulsation.} 
\label{Table:PulsationPropertiesPopII}
\begin{tabular}[t]{lcrrc}\toprule
	 \centering
	  ($M_{\mathrm{ZAMS}},M_{\mathrm{final}}$) in s.u. 
	               &  $\log \llsol$ & P / d   & $\#$ of puls. cycles & {\tt Blocker\_scaling\_factor} \\ 
	 \midrule
	  (3.0, 0.982) & 4.4  &  1097    &   3 &  0.05 \\ 
	 \bottomrule
\end{tabular}\par
} 
\end{table}

\smallskip

To reach sufficiently high $L/M$ ratios for pulsations to unfold, the
lowest-mass model sequences, $(1.5, 0.628)$ for PopI and $(3.0,0.982)$
for PopII composition, had both to be in the phase of a He-shell
flash.  Hence, at the low-mass end, the chances for pulsations to
occur are low because the respective AGB star must be in the proper,
comparatively short-lived phase of the ThP cycle during which the
He-shell is thermally unstable.  Going up in mass, pulsations can
however develop essentially at any phase through a ThP cycle. In the
higher-mass PopI model sequences computed in this study, the H- and
He-shell were both active, with the H-shell being the dominant energy
source, when the pulsations developed.

\bigskip
\centerline{\afgsection{4. THE NATURE OF THE PULSATIONS}}

\begin{wrapfigure}{r}{0.50\textwidth}
	  \vspace{-20pt}
	  \begin{center}{
			\includegraphics[width=0.49\textwidth]{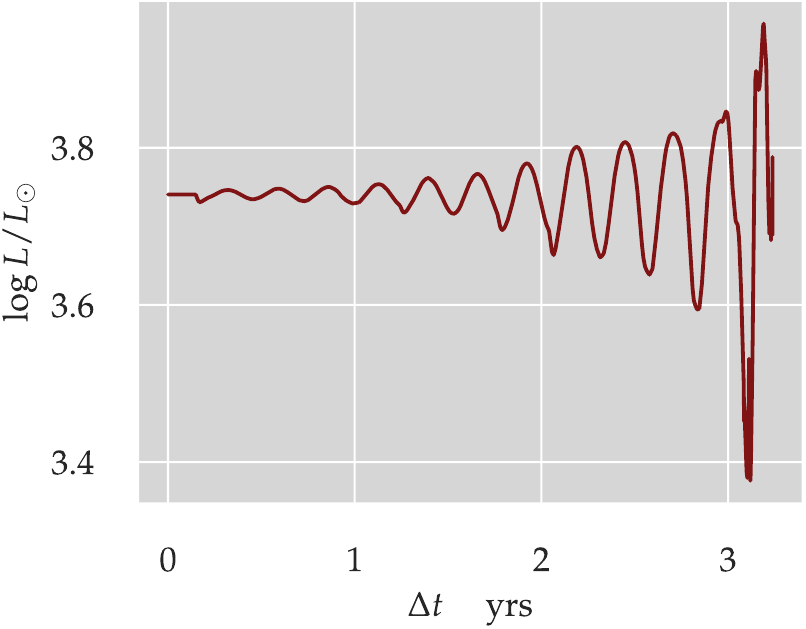}	
	  }\end{center}
	  \vspace{-15pt}
	  \caption{The temporal growth of the luminosity variation
	  		   during the terminal pulsational instability of 
	  		   the $(2.0,0.570)$ sequence. 
	  	      }\label{fig:M020Z02_Lightvar}
	 \vspace{-20pt}	
\end{wrapfigure}
This section focuses on the pulsation that develops in the PopI $(2.0,
0.570)$ sequence. It is the one whose terminal pulsation period is
comparatively short with about $98$~days. The astrophysical
variability behavior is nonetheless representative, it does not differ
much from the longer-period pulsations of the sample presented
here. The relatively large number of pulsation cycles that could be
followed numerically invite some standard diagnostics from classical
pulsation theory to be applied.

Figure~\ref{fig:M020Z02_Lightvar} illustrates the rapid growth of the
amplitude of the cyclic luminosity variation; the zero-point of the
abscissa was chosen arbitrarily around the onset of the
variability. Apparently, the pulsation does not grow out of
computational noise but gets initiated by a small but finite
amplitude-kick, which is likely picked up during the numerical
relaxation process.

\begin{wrapfigure}{r}{0.50\textwidth}
	  \vspace{-20pt}
	  \begin{center}{
			\includegraphics[width=0.49\textwidth]{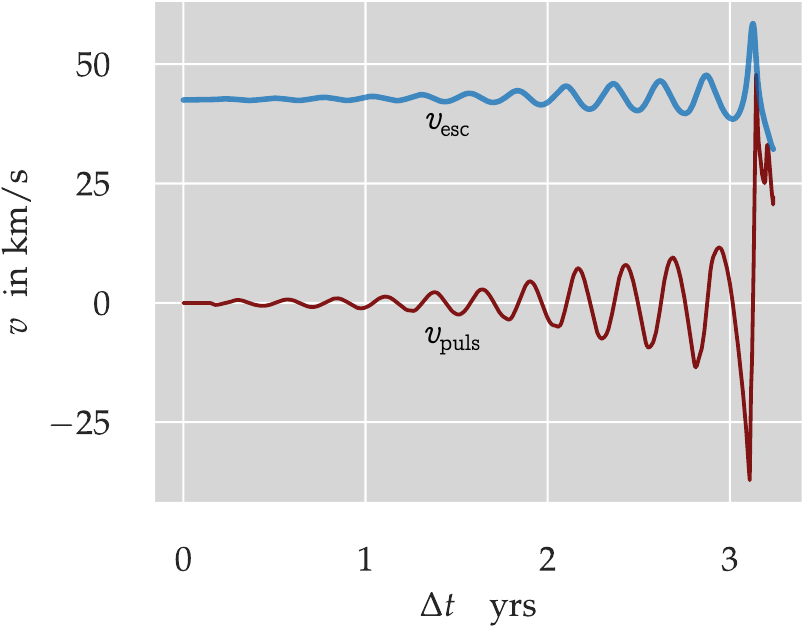}	
	  }\end{center}
	  \vspace{-15pt}
	  \caption{After less than a dozen pulsation cycles the
	           pulsational radial velocity at the
	  		   surface approaches escape speed.  
	  	      }\label{fig:M020Z02_velocities}
	 \vspace{-10pt}	
\end{wrapfigure}

The pulsational radial velocity ($v_{\mathrm{puls}}$) of the star's
surface (Fig.~\ref{fig:M020Z02_velocities}) visualizes once again the
rapid growth of the pulsation; the nonlinearity is well expressed
after about six pulsation cycles.  This development goes in par with
the bolometric light change displayed in
Fig.~\ref{fig:M020Z02_Lightvar}. The comparison of $v_{\mathrm{puls}}$
with the escape velocity from the surface ($v_{\mathrm{esc}}$)
explains why \MESA\,eventually fails to converge or stalls once the
outermost layers approach escape velocity. It is a common feature of
\emph{all }the pulsating model sequences in
Table~\ref{Table:PulsationProperties} that their photospheric
pulsation velocity approach escape velocity at the end of the
simulation runs.

\begin{wrapfigure}{l}{0.53\textwidth}
	  \vspace{-25pt}
	  \begin{center}{
			\includegraphics[width=0.52\textwidth]{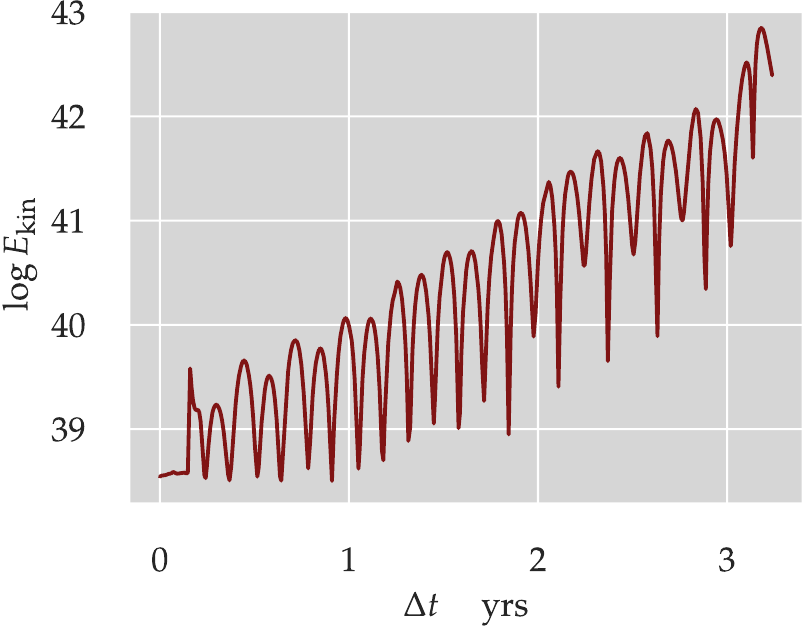}	
	  }\end{center}
	  \vspace{-15pt}
	  \caption{On a logarithmic scale, the growth of the star's 
	           kinetic energy (measured in erg) is essentially linear.   
	  	      }\label{fig:M020Z02_Ekin}
	 \vspace{-10pt}	
\end{wrapfigure}

The pulsational instability depicted in
Figs.~\ref{fig:M020Z02_Lightvar} and \ref{fig:M020Z02_velocities}
gives the impression that it grows exponentially. This is indeed the
case as can be directly deduced from a plot of the temporal evolution
of the model star's directed kinetic energy.  The appropriate
quantity, $E_{\mathrm{kin}}$, is plotted in
Fig.~\ref{fig:M020Z02_Ekin} on a logarithmic scale. A decently
matching linear fit to $E_{\mathrm{kin}}(\Delta t)$ yields an
e-folding time of about 140 days. This means that the kinetic energy
grows by a factor e in less than two pulsation cycles. According to
Fig.~\ref{fig:M020Z02_Ekin}, the directed kinetic energy starts at
around $5\cdot10^{38}$~erg and quickly grows to about
$6\cdot10^{42}$~erg.  Within less than a dozen cycles,
$E_{\mathrm{kin}}$ grows by about four orders of
magnitude. Nonetheless, the total (potential and internal) energy of
the star is of the order of $-10^{49}$~erg. Hence, the kinetic energy
of the pulsation remains, even at maximum amplitude, a minor
contribution to the energy budget of the star. From the energetic
point of view, the initial growth of the coherent pulsational motion
with its $5\cdot10^{38}$~erg is~--~numerically seen~--~still
developing out of the noise.

\begin{figure}
  \begin{center}
     \includegraphics[width=0.98\textwidth]{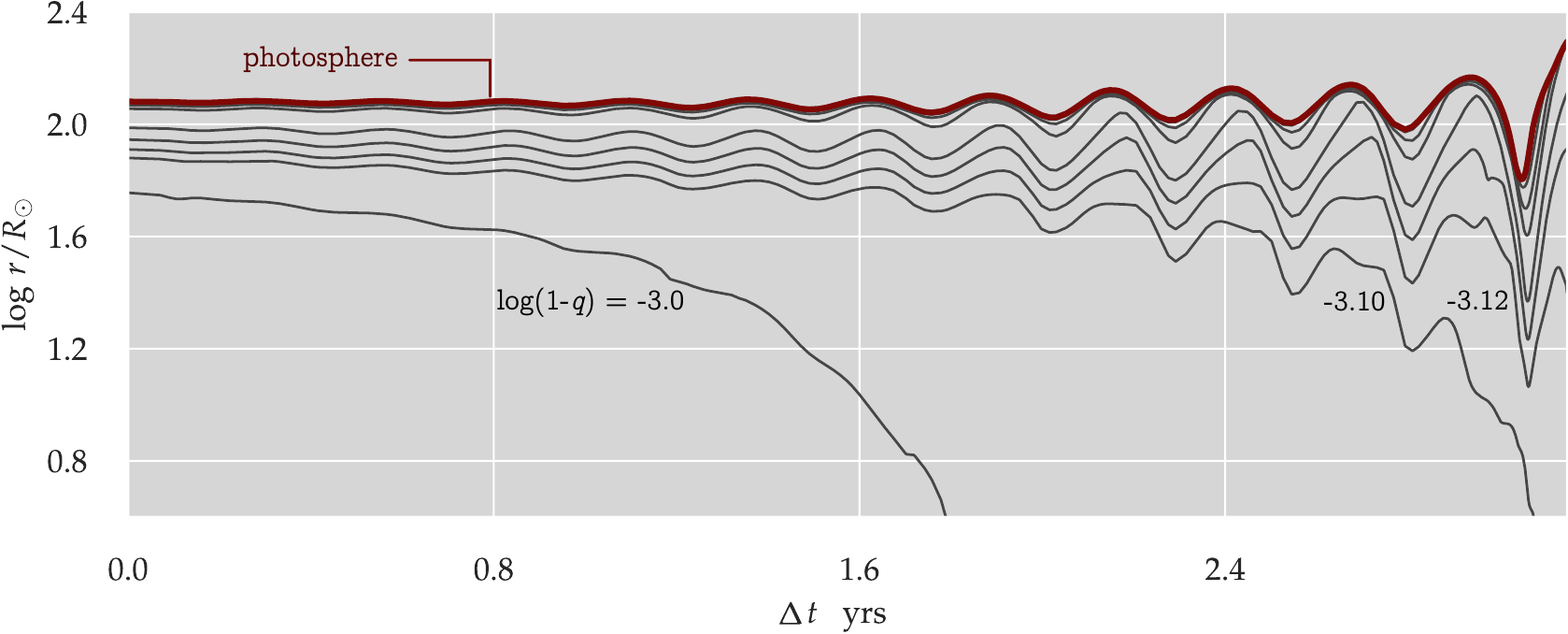}
  \end{center}
  \vspace{-15pt}
  \caption{Temporal radius variation of selected mass shells 
           close to the star's photosphere. Only the very
           outermost layers participate in the cyclic variability.
          }\label{fig:M020Z02_puls_radii}
\end{figure}

The temporal variation over the almost 12 pulsation cycles of the 
photospheric radius (red) and radii of selected mass levels close 
to the star's surface are assembled in Fig.~\ref{fig:M020Z02_puls_radii}. 
The deepest-lying mass level that is 
plotted lies at $\log(1-m/\mast) =-3.0$; i.e. only the outermost 
about $10^{-3}$ of the star's mass is contained in the plot. 
The various mass shells were chosen purely for illustrative purposes. 
To quantify the unusual behavior of the mass layers in the plot, 
two more of them are labeled numerically.

The $\log(1-m/\mast) =-3.0$ mass-level is hardly affected by the
pulsation. Even though it shows some cyclic modulation it contracts
monotonically as does the majority of the star already.  The
mass-level tagged with $-3.10$ participates in the pulsation almost to
the end of the simulation but eventually joins the contraction of the
rest of the model star.  Mass-layers above the $-3.12$ level, on the
other hand, constitute a very thin envelope that oscillates to the
very end of the computation and whose compression- and
expansion-kinematics hints at the formation of shock fronts during
contraction.  The picture that emerges from
Fig.~\ref{fig:M020Z02_puls_radii} is that of a mostly contracting
interior plus a thin pulsating outer envelope that evolves towards
detachment on a time-scale of years or less. These two regions clamp a
low-density cavitation-like layer that induces temporarily pronounced
density inversions at its outer edge.

During cycle~10, the second to the last cycle shown in
Fig.~\ref{fig:M020Z02_puls_radii}, the amplitudes of the
pulsation-induced variations in the outer envelope 
are considerable and nonlinearities are already well developed. 
Figure~\ref{fig:M020Z02_drr0_profiles_cycl10} illustrates 
from a complementary angle the relative radius variability
contained in Fig.~\ref{fig:M020Z02_puls_radii}.
\textbf{\begin{figure}[h!]
\centering
\begin{minipage}{.48\textwidth}
  \centering
  \includegraphics[width=.98\linewidth]
                  {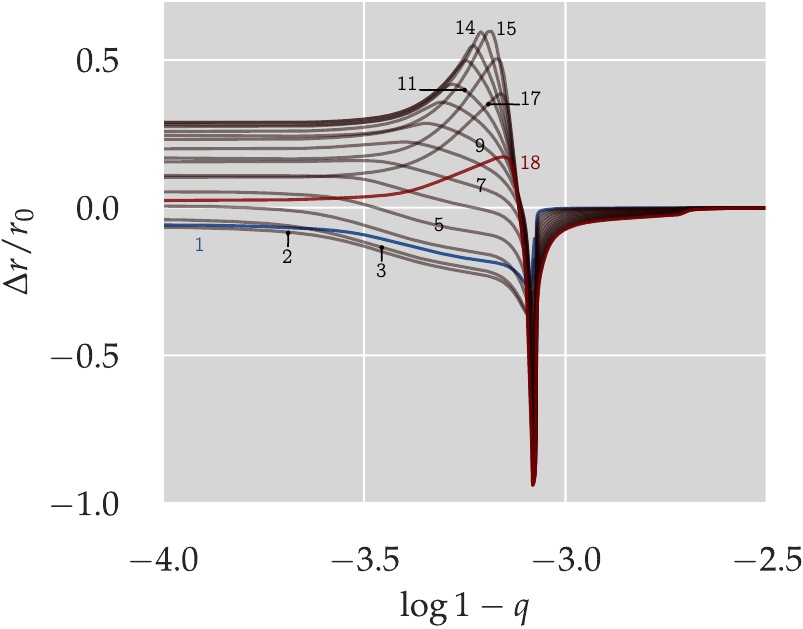}
  \captionof{figure}{Relative radius-variation profiles
  	                 at 18 epochs during pulsation-cycle 10.
                    }
  \label{fig:M020Z02_drr0_profiles_cycl10}
\end{minipage}%
\hfill
\begin{minipage}{.48\textwidth}
  \centering
  \includegraphics[width=.98\linewidth]
                  {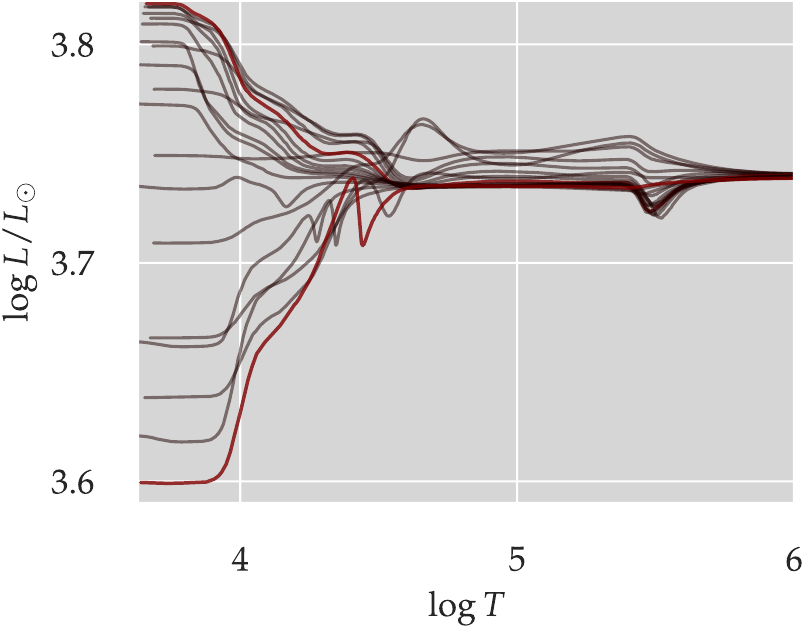}
  \captionof{figure}{At the same phases as in
                     Fig.~\ref{fig:M020Z02_drr0_profiles_cycl10},         
                     a collection of luminosity profiles.}
  \label{fig:M020Z02_lum_profiles_cycl10}
\end{minipage}
\end{figure}}

\MESA\,computed 18 epochs to cover pulsation-cycle 10. The arbitrary
chosen epoch 0 serves as the reference profile relative to which the
quantity $\Delta r / r_0 \equiv (r - r_0)/ r_0$ is calculated.  To
help to read the progression of the relative radius change over a
pulsation cycle the profiles in
Fig.~\ref{fig:M020Z02_drr0_profiles_cycl10} are numbered.  The first
epoch is additionally colored in blue, the last epoch of the chosen
cycle is plotted in red. This coloring helps to emphasize the temporal
radius behavior in the region $\log (1-q) \gtrsim -3.1$: As pointed
out in Fig.~\ref{fig:M020Z02_puls_radii} already, these deeper layers
do not participate in the oscillatory motion but only shrink
monotonously. The contraction magnitude decreases with mass depth and
vanishes before the outer edge of the H-burning shell is reached.  All
in all, the relative radius-variation over a cycle of our pulsating
early~--~post-AGB stars does not resemble the eigenfunctions one is
used to from classical pulsators.

The variation of the spatial luminosity profiles over pulsation cycle
10 is a further illustration of how different the pulsations of the
early~--~post-AGB stars are from classical pulsators.  In accordance
with Fig.~\ref{fig:M020Z02_puls_radii}, the pulsation affects only the
regions cooler than $10^6$~K.  Between the $Z$-bump and the
He$^+$-partial ionization zone ($\log T \approx 4.5$), on the average
the luminosity rises only; i.e.  it is not an oscillation about a
quasi-equilibrium configuration.  This changes in regions closer to
the photosphere, there the luminosity varies such that it falls below
the luminosity prevailing in the pulsation-unaffected depth of the
star ($\log T \gtrsim 6$) at some phases and exceeds that luminosity
about half a cycle later. Except for the regions with $\log T \lesssim
4.5$, the spatial variability is unlike the harmonic behavior seen in
the common pulsating stars.  To emphasize the peculiarity of the
variability profiles, maximum and minimum light phases are highlighted
in red in Fig.~\ref{fig:M020Z02_lum_profiles_cycl10}.

\begin{wrapfigure}{r}{0.53\textwidth}
	  \vspace{-20pt}
	  \begin{center}{
			\includegraphics[width=0.52\textwidth]{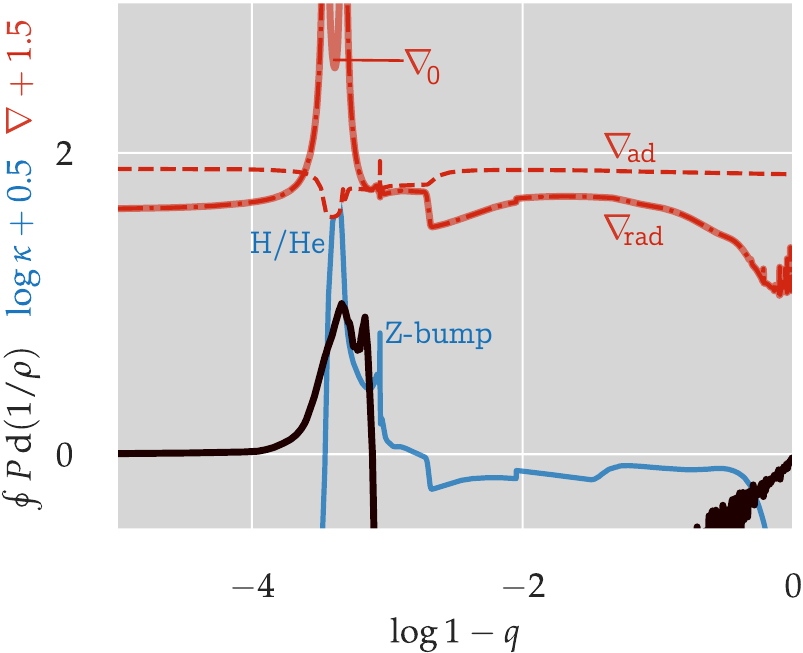}	
	  }\end{center}
	  \vspace{-15pt}
	  \caption{Differential work over a pulsation cycle (black) and
	  	       the spatial run of the vertically shifted 
	  	       Rosseland opacity of the
	  	       first model of the cycle (blue) to help to 
	  	       identify the source of pulsation driving.  
	  	      }\label{fig:M020Z02_PdV}
	 \vspace{-20pt}	
\end{wrapfigure}
None of the pulsations discussed here could ever be tracked
to saturation, i.e. they never reached strict periodicity. 
Nonetheless, driving and damping regions in the envelopes were tried 
to be identified by computing the differential work integral 
\[
\Delta W =       \oint\displaylimits_{\mathrm{cylce}}\! P\,\diff_t \Delta V \diff t
         \propto \oint\displaylimits_{\mathrm{cylce}}\! P\,\diff 
                                              \left(\frac{1}{\rho}\right)\,,
\]
per mass shell $\Delta m$ over a pulsation cycle. The differential
work integrals were computed during the still rather smooth cycles~8
and~9.  Different initial phases, from which a quasi-cycle was
defined, were tried out.  The results all looked like what is plotted
in Fig.~\ref{fig:M020Z02_PdV}, where the differential work is
arbitrarily normalized to maximum driving.  Even though the details of
the driving region ($\Delta W > 0$) varied depending on the starting
phases of a cycle, the driving and damping region remained spatially
stable, independent of the particular numerical choices. To correlate
the driving region with envelope properties,
Fig.~\ref{fig:M020Z02_PdV} shows also the vertically shifted
Rosseland-opacity profile of the starting model of the cycle. All the
pulsation driving is confined to the partial ionization-region of
H/He. At the base of the $Z$-bump, the pulsation is already strongly
damped. The damping computed in the even deeper, i.e. hotter layers
becomes eventually useless because the respective mass shells do not
participate in the cyclic pulsation; they undergo purely monotonous
state changes only (cf. Fig.~\ref{fig:M020Z02_puls_radii}) so that the
path integral along a \emph{cycle }has no basis anymore.  The
important take-away is that driving occurs in the partial
ionization-regions of H and He. These regions coincide with the
large-amplitude, also spatially pulsational variability seen in the
profiles of the model stars.

\medskip

\begin{wrapfigure}{l}{0.51\textwidth}
	  \vspace{-25pt}
	  \begin{center}{
			\includegraphics[width=0.50\textwidth]{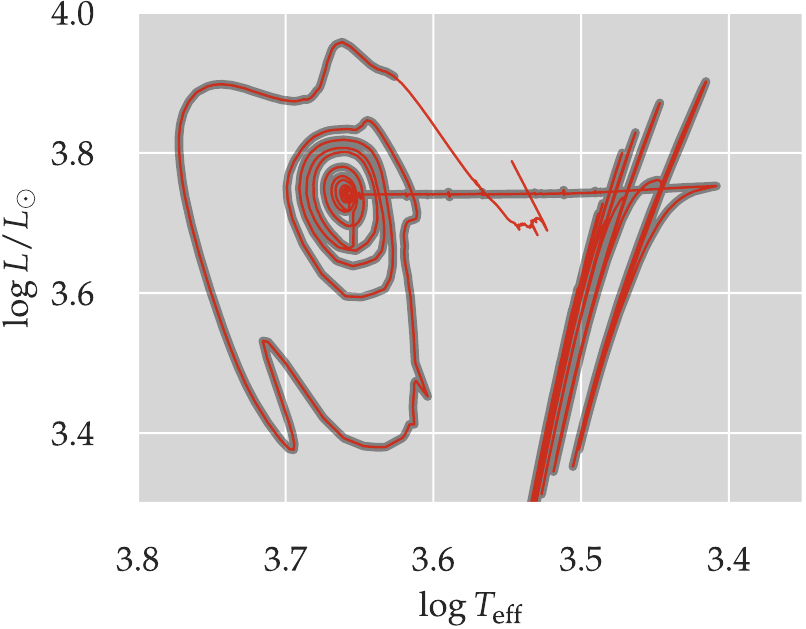}	
	  }\end{center}
	  \vspace{-15pt}
	  \caption{ Terminal evolution along the AGB of 
	  	    the (2.0, 0.570) sequence, once computed dynamical~(red) 
	  	    and once in QHE (thick grey line).
	  	      }\label{fig:M020Z02_DynVsQHE}
	 \vspace{-10pt}	
\end{wrapfigure}

We finish with the successful recovering the pulsations also in the
QHE ansatz.  A few cases only were computed so far. Particularly
interesting are the $\mZAMS = 2\,\msol$ cases: There, \emph{the QHE
and dynamical evolution computations behave identical}, as exemplified
by the (2.0, 0.570) sequence shown in
Figure~\ref{fig:M020Z02_DynVsQHE} where the terminal evolutionary
tracks on the AGB and the departure therefrom are mapped onto the HR
plane.  The result of the dynamical treatment is traced out by the red
line.  The broader grey line shows the evolution of the same model
star in QHE mode.  Even the small bumps along the evolutionary tracks
are identical.  The two $2\,\msol$ cases are important counterexamples
to the interpretation of the pulsations presented here as radial
acoustic modes. Thanks to the pulsations persisting also in the QHE
approximation it is clear that these pulsations are \emph{not }excited
acoustic waves as known from classical pulsators. Furthermore, the
dynamical treatment of the evolution problem does not lead to a more
dynamic or even monotonic instability.
 
Under the physical conditions prevailing in the envelopes of very
luminous AGB stars, thermal and dynamical timescales become comparable
and the distinction between acoustic and thermal modes gets ever more
difficult to maintain. It has been known for a long time
\citep[e.g.][]{Wood1976, gagla90b, SBG98} that \emph{strange modes
}that have no counterparts in the adiabatic acoustic mode-spectrum of
a star might develop around $L/M \sim \mathcal{O}(10^4)$ with
pulsation periods that can exceed those of the fundamental radial
p-mode. Such modes can go unstable with very large growth rates. These
properties are all features which also fit in the pulsational
instability encountered in the early~--~post-AGB model stars.

\bigskip

\centerline{\afgsection{5. WRAPPING IT ALL UP}}

Lagrangian, dynamical \MESA\,evolution computations revealed, under
suitably favorable conditions, the development of rapidly growing
radial pulsations shortly after or around the termination of the AGB
evolution of initially low- and intermediate mass stars.  The periods
found range from about 100 to over 1000~days.  The e-folding times of
a few periods' length are much shorter than what is encountered in
classical pulsators.  The behavior is reminiscent of the pulsations of
massive red-supergiants \citep[e.g.][]{Yoon2010} that were also
successfully tracked with \MESA\,\citep{Paxton2013}.  With the model
sequences discussed in this proof-of-concept report, dynamical
\MESA\,computations demonstrate that they are capable of recovering
oscillatory instabilities at the end of AGB evolution of low-mass
stars as reported by \citet{Wagenhuber1994}. The latter authors found
such pulsations in their QHE computations confined to around maximum
instability of the models' He-shell.  Within a few years of a star's
evolution time, i.e. after only a few pulsation cycles the model stars
build up superficial radial pulsation velocities that approach escape
speed.  Eventually, convergence of the models failed due to the
constraints of the numerical approach to evolve these stars.

In two \MESA\,model sequence so far, the QHE and the dynamical
evolution of the model stars are found to behave identical.  Hence,
this step beyond what \citet{Wagenhuber1994} found does not
substantiate their expectation that the instability seen in their QHE
treatment would turn into a dynamical instability.  It is rather more
obvious to understand these pulsations as thermal strange modes, seen
directly in their nonlinear regime.

The pulsations are found to develop in a very
thin~--~$\mathcal{O}(\lesssim 10^{-3}\,\msol)$~--~surface layer whose
oscillation amplitude grows rapidly on top of a monotonously shrinking
interior. Between these two regions a low-density cavitation-like
region (cf. Fig.~\ref{fig:M020Z02_puls_radii}) builds up.  At the low
densities and the strong radiation fields that prevail in the
outermost layers of the pulsating post-AGB stars, matter and radiation
are likely out of equilibrium so that a radiation-hydrodynamic
treatment of the problem is called for. More appropriate computational
tools that can follow the dynamics also across the transition from the
stellar surface into the circumstellar medium must scrutinize the
pulsations found in the \MESA\,computations. It will be interesting to
learn if the pulsations persist and if they do, what the dynamics of
the density-inversion region (the cavitation) is, and how much mass
the pulsation is able to shed.

Even in the coolest early~--~post-AGB pulsators
(e.g.~Fig.~\ref{fig:M020Z02_LtoM}) the envelope convection zones are
shallow. Mostly, the regions around the opacity $Z$-bump and the H/He
partial ionization zones are convectively unstable.  The way
mixing-length convection is treated in stellar evolution computations
means that convection adapts instantaneously to the local physical
states through the star.  The encountered pulsation periods of
hundreds of days are, however, not that different from the convection
timescale.  Figure~\ref{fig:M020Z02_PdV} shows representatively (in
red) the pertinent temperature gradients in early~--~post-AGB
pulsators.  The dashed line traces $\nabad$, $\nabrad$ is shown as a
dot-dashed line, which is mostly covered by the thicker, transparent
full line of $\nabla_0$, the effectively prevailing temperature
gradient. The very thin (in mass) convection zone at the $Z$-bump is
still adiabatic. In contrast, the convection region around H/He
partial ionization, where pulsation driving occurs, is very
non-adiabatic. Hence, only a fraction of the total flux is effectively
transported by material motion in the driving region. As it was
already called for by \citet{Wagenhuber1994} under comparable
circumstances, the effect of time-dependent convection on the
instability needs to be investigated. Such an endeavor should be
feasible now after the local, time-dependent convection formalism of
Kuhfu\ss\,was added to \MESA's toolbox \citep{Jermyn2023}.

Based on the presented results, early~--~Post-AGB pulsations can
emerge over a broad range of phases of a ThP cycle.  The rapid growth
of the instability has the potential to shed much of the superficial
layers~--~of the order of $10^{-4} - 10^{-3}\,\msol$~--~within a few
years.  It should be interesting to learn what fraction of the
observed $25\%$ H-poor central stars of planetary nebulae could be
produced by first-time descendants from the AGB already, i.e.  by
pulsating early~--~post-AGB stars. If this is possible, this should
induce a mass-dependent signature. For lower-$\mZAMS$ stars to develop
pulsations a He-shell instability seems to be necessary . Hence, this
scenario would naturally satisfy the requirement of a correlation of
mass-loss and thermal-pulse cycle \citep{Bloecker2001}.  Higher
initial masses, on the other hand, are seen to leave the AGB and
develop pulsations almost at any phase during the ThP
cycle. Furthermore, a metallicity dependence is to be expected too.
Metal-poorer stars were found to develop early~--~post-AGB pulsations
for comparatively higher $\mZAMS$ only.

Post-AGB stars that live through a (very) late thermal pulse
\citep{Bloecker2001} can return to the AGB as so-called born-again AGB
stars and reach thereby sufficiently high luminosities and low
effective temperatures to (possibly once again) develop pulsations of
the kind presented in this study. The peculiar variable stars FG Sge
\citep{Jurcsik1999}, V~605~Aql, and Sakurai's object
\citep{Lawlor2002} might be examples of or candidates for such
objects.  During such a very late phase of recurring dynamical
mass-loss, any remaining thin H-rich envelope might become either very
depleted or even get completely lost.

This report raises more questions than it answers. Nevertheless, a
selfconsistent mechanism~--~computationally established with complete
stellar-evolution models~--~is reiterated on to argue that it might
serve as the physical machinery that drives the hitherto hypothetical
superwind.  The preliminary results computed with \MESA\,hint at
several promising and potentially rewarding research avenues in the
field of AGB and post-AGB stellar astrophysics that should be
accessible now to the available computational tools.

\bigskip

\textsc{ACKNOWLEDGEMENTS: }This work relied substantially on NASA's
Astrophysics Data System. The model stars were computed
with the \MESA\,~software-instrument, 
version r$21.12.1$ \citep{Paxton2019}.
Postprocessing \MESA{\tt star}\,models and plotting 
benefited from Warrick Ball's 
\href{https://github.com/warrickball/tomso}{\texttt{tomso}},
\texttt{numpy }\citep{harris2020array},
\texttt{scipy }\citep{2020SciPy}, and
\texttt{matplotlib }\citep{Hunter2007}, respectively. I am 
profoundly grateful to have had free access to all these 
software tools and to the full texts of much of the cited
scientific literature.

\bigskip\bigskip
\centerline{\afgsection{APPENDIX}}
\begin{verbatim}
&star_job
      !...Nuclear physics
      change_net    = .true.     
      new_net_name  = 'approx21.net' 

      ! Macrophysics part
      change_v_flag = .true. 
      new_v_flag    = .true.
/ !end of star_job namelist

&eos
     ! eos options
     ! see eos/defaults/eos.defaults
/ ! end of eos namelist

&kap
      kap_file_prefix      = 'a09'
      kap_lowT_prefix      = 'lowT_fa05_a09p'
      kap_CO_prefix        = 'a09_co'
      use_Type2_opacities  = .true.   
      Zbase                = 0.02d0             
/ ! end of kap namelist

&controls
      !...to start evolution from ZAMS 
      initial_mass = 1.5d0      
      initial_z    = 0.02d0  

      energy_eqn_option   = 'dedt'
      use_gold_tolerances = .true.  

      !...Convection
      mixing_length_alpha = 1.8d0
      MLT_option          = 'Cox' 

      use_Ledoux_criterion  = .false.
      do_conv_premix        = .false. 

      !...Overshooting
      overshoot_scheme(1)    = 'exponential'
      overshoot_zone_type(1) = 'burn_H'
      overshoot_zone_loc(1)  = 'core'
      overshoot_bdy_loc(1)   = 'top'
      overshoot_f(1)         = 0.012
      overshoot_f0(1)        = 0.002

      overshoot_scheme(2)    = 'exponential'
      overshoot_zone_type(2) = 'nonburn'
      overshoot_zone_loc(2)  = 'shell'
      overshoot_bdy_loc(2)   = 'bottom'
      overshoot_f(2)         = 0.022
      overshoot_f0(2)        = 0.002

      make_gradr_sticky_in_solver_iters = .true.

      !...Mass-loss processes
      cool_wind_RGB_scheme   = 'Reimers'
      Reimers_scaling_factor = 0.1        
      cool_wind_AGB_scheme   = 'Blocker'
      Blocker_scaling_factor = 0.2         
      RGB_to_AGB_wind_switch = 1d-4   	  

      !...time-step & grid control
      delta_HR_limit    = 0.02   
      
/ ! end of controls namelist
\end{verbatim}

\bibliographystyle{aa}
\bibliography{StarBase}

\end{document}